\documentclass{PoS}

\title{The First Station of the Long Wavelength Array}

\ShortTitle{The First Station of the Long Wavelength Array}

\author{\speaker{Patricia Henning}\\
        University of New Mexico\\
        E-mail: \email{henning@cosmos.phys.unm.edu}}

\author{Steven W.~Ellingson\\
        Virginia Polytechnic Institute and State University\\
        E-mail: \email{ellingson@vt.edu}}

\author{Gregory B.~Taylor, Joseph Craig, Ylva Pihlstr\"om, Lee J Rickard\\
        University of New Mexico\\
        E-mail: \email{gbtaylor@unm.edu}, \email{joecraig@unm.edu}, \email{ylva@unm.edu}, \email{lrickard@unm.edu}}

\author{Tracy E.~Clarke, Namir E.~Kassim\\
       U.S. Naval Research Laboratory\\ 
       E-mail: \email{tracy.clarke@nrl.navy.mil}, \email{Namir.Kassim@nrl.navy.mil}}

\author{Aaron Cohen\\
        U.S. Naval Research Laboratory and Johns Hopkins Applied Physics Laboratory\\
        E-mail: \email{acohen00@gmail.com}}

\abstract{The Long Wavelength Array (LWA) will be a new multi-purpose radio telescope 
operating in the frequency range 10--88 MHz. 
Upon completion, LWA will consist of 53 phased array ``stations'' distributed 
over a region about 400~km in diameter in the state of New Mexico. Each station 
will consist of 256 pairs of dipole-type antennas whose signals are formed into 
beams, with outputs transported to a central location for high-resolution
aperture synthesis imaging.  The resulting image sensitivity is estimated to be a 
few mJy ($5\sigma$, 8~MHz, 2 polarizations, 1 hr, zenith) in 20--80~MHz; with
resolution and field of view of ($8''$,$8^{\circ}$) and ($2''$,$2^{\circ}$) at 20~MHz 
and 80~MHz, respectively. 
All 256 dipole antennas are in place for the first station of the LWA (called LWA-1), and commissioning
activities are well underway.  
The station is located near the core of the EVLA, and is expected to be fully operational in early 2011.}

\FullConference{ISKAF2010 Science Meeting \\
		 June 10 -14 2010\\
		 Assen, the Netherlands}

\begin{document}

\section{The Long Wavelength Array}

We are living in an era of resurgent interest in astronomy at long wavelengths,
as witnessed by the development of new low frequency telescopes around the world,
including LOFAR, MWA, PAPER, all discussed in this volume, and the LWA.

The LWA is designed for both long-wavelength astrophysics and ionospheric 
science.  
Science to be addressed by the LWA includes cosmic evolution, the acceleration 
of relativistic particles, physics of the interstellar and intergalactic media, 
solar science and space weather, and ``discovery science''; that is, the search 
for previously unknown sources and phenomena \cite{NEK-ASP345}.  
Specific objectives for LWA are spelled out in \cite{LWA117}.

\begin{figure}
\begin{center}
\includegraphics[width=.8\textwidth]{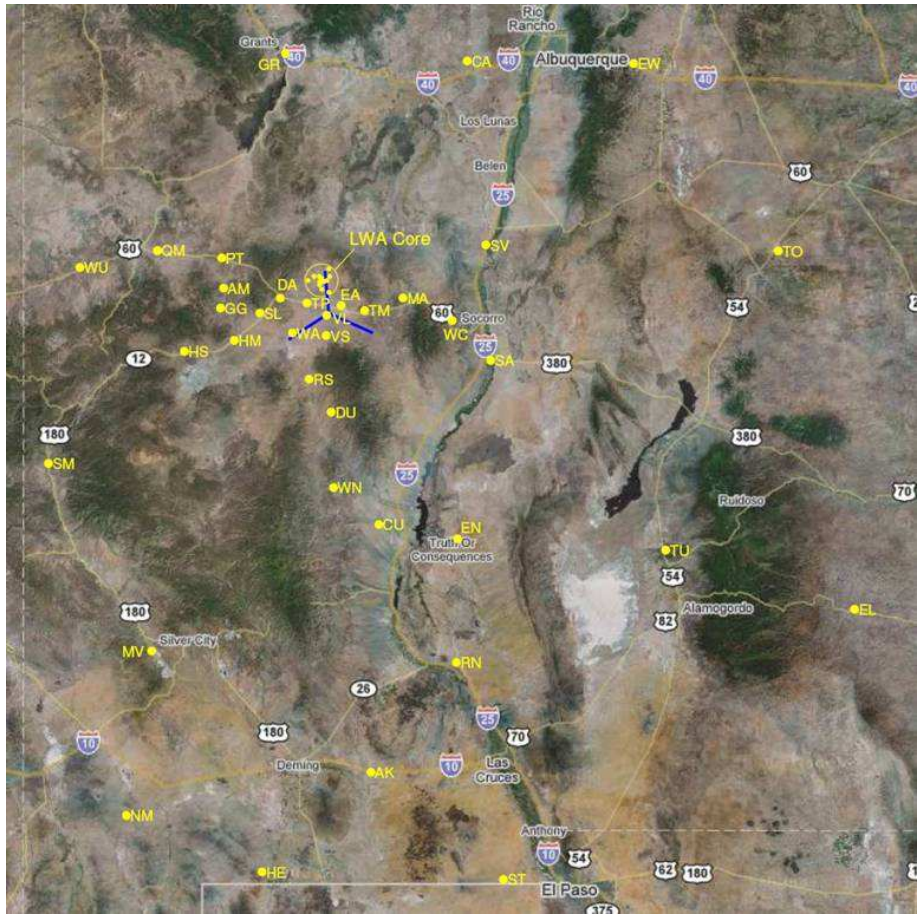}
\end{center}
\caption{Planned LWA station locations across the state of New Mexico.
The ``Y'' of the EVLA is shown in blue.}
\label{stations}
\end{figure}

Upon completion, LWA will consist of 53 electronically-steered phased array 
``stations,'' each consisting of 256 pairs of dipole-like antennas, operating 
with Galactic noise-limited sensitivity over the frequency range 20--80~MHz, with reduced sensitivity over 10--88~MHz. 
The stations will be distributed over the state of New Mexico, 
(Figure~\ref{stations}), with maximum baselines of up to 400~km,  
yielding resolution of $8''$ and $2''$ at 20~
MHz and 80~MHz respectively.
%Stations will be capable of operating as independent radio telescopes; however 
%normally beams formed by the stations will be transmitted to a central location 
%and correlated to form images using aperture synthesis techniques \cite{TMS01}. 
Beams formed by the stations will be transmitted to a central location and 
correlated to form images using aperture synthesis techniques.
The full array is expected to reach mJy-class sensitivity. 
%\cite{TMS01}.  
Stations will also be capable of operating as independent radio telescopes.

\section{LWA-1, the First Station of the LWA}

\begin{figure}
\begin{center}
\includegraphics[width=.8\textwidth]{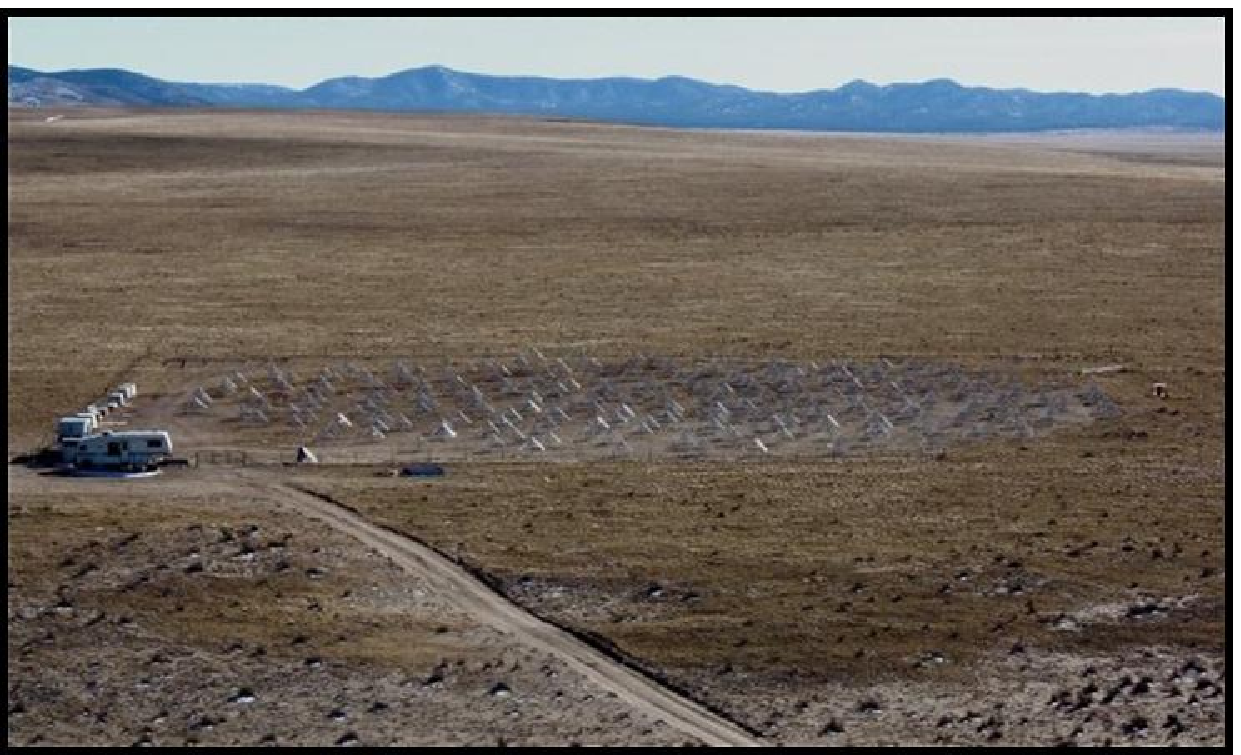}
\end{center}
\caption{The first station of the LWA, as seen from the EVLA antenna assembly building.
The equipment shelter is seen to the left.}
\label{lwa1}
\end{figure}

\begin{figure}
\begin{center}
\includegraphics[width=.9\textwidth]{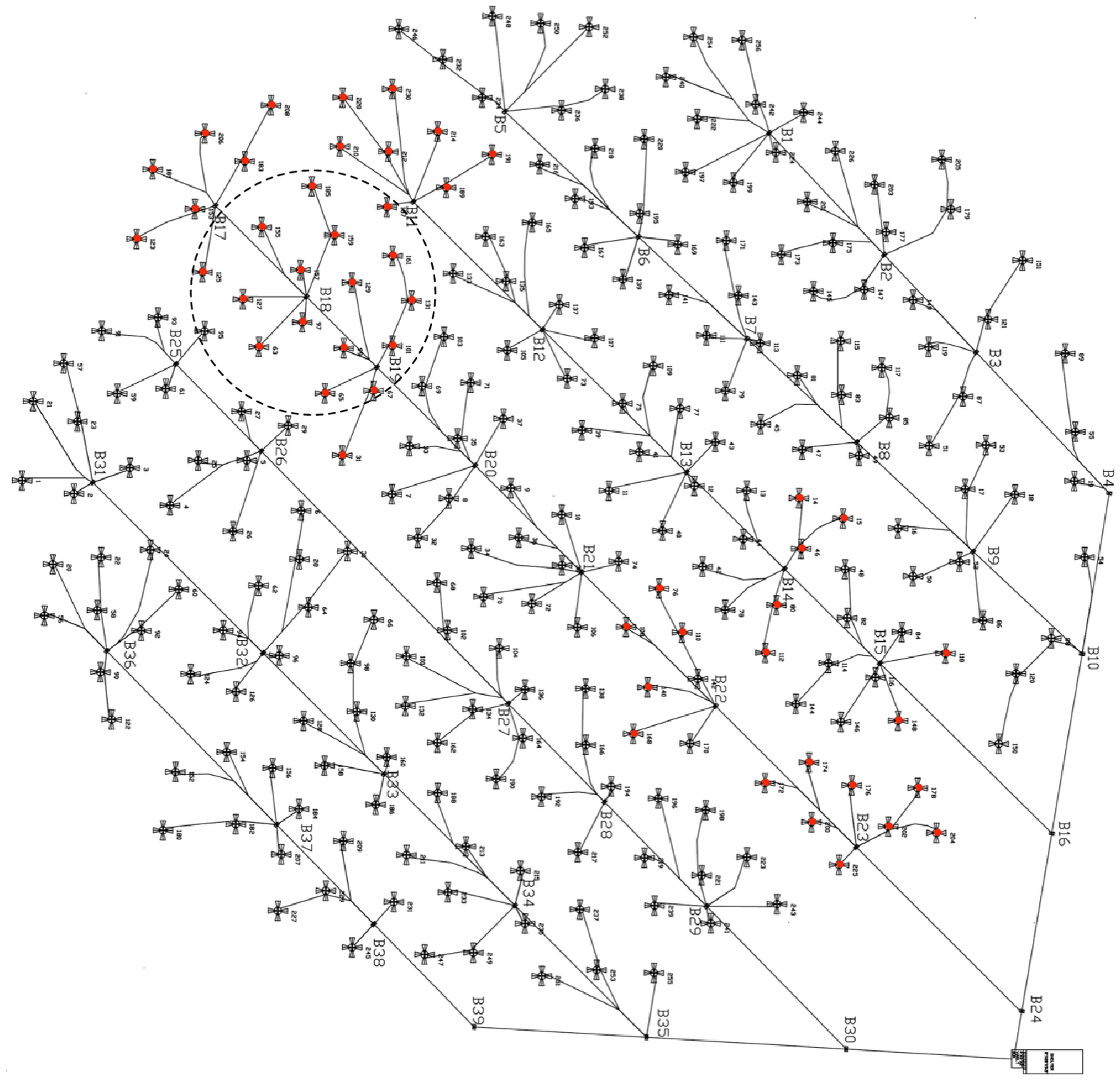}
\end{center}
\caption{Diagram of antenna locations and cabling in LWA-1.  
The circle indicates antennas which participated in early tests.}
\label{layout}
\end{figure}

The first station of the LWA, called LWA-1, is being constructed near
the core of the EVLA.  Its 256 crossed dipole antennas (shown in 
Figure~\ref{lwa1}) were completed
in November 2009.  The elements are distributed in a
$\sim$100-m aperture, producing beam FOV of $8\deg$ and $2\deg$ at
20~MHz and 80~MHz.  Every element is digitized to allow independent
pointings of beams, and all-sky snapshot imaging.  Using 256
dual-polarization antennas results of spacings of 3 x Nyquist at 80
MHz.  The pseudorandom distribution of the antennas mitigates against aliasing
(Figure~\ref{layout}).  The minimum separation between antennas of 5~m
allows easy access for maintenance, and also reduces beam
desensitization due to sky noise correlation \cite{SWG-IEEE97}.
Various ionospheric, solar, and especially Galactic science goals
require the ability to observe towards declinations which appear low
in the southern sky from New Mexico.  To compensate for the
elevation-plane widening of the beam for these observations, the
station footprint has been made somewhat elliptical with a NS:EW axial
ratio of 1.1.

\subsection{RF Signal Path, and System Status}

The signal from every antenna is processed by a dedicated
direct-sampling receiver consisting of an analog receiver (ARX) and an
analog-to-digital converter (A/D) which samples 196 million samples
per second (MSPS).  Beams are formed using a time-domain delay-and-sum
architecture, which allows the entire 10--88~MHz passband associated
with each antenna to be processed as single wideband data stream.
Eventually beams from the stations will be correlated in a central
location, but for individual stations it is also possible to record the beam
data for later processing.  To facilitate commissioning activities,
diagnostics, and certain types of science observations requiring
all-sky field-of-view (FOV), the station electronics will also have the capability to
coherently capture and record the output of all A/Ds, where each A/D
corresponds to one antenna.  This will occur in two modes: the
``transient buffer -- wideband'' (TBW) allows the raw output of the
A/Ds to be collected continuously, but only for $\sim100$~ms at a
time.  The ``transient buffer -- narrowband'' (TBN), in contrast,
allows a single tuning of $\sim100$~kHz bandwidth to be recorded
indefinitely.

\begin{figure}
\begin{center}
\includegraphics[width=.8\textwidth]{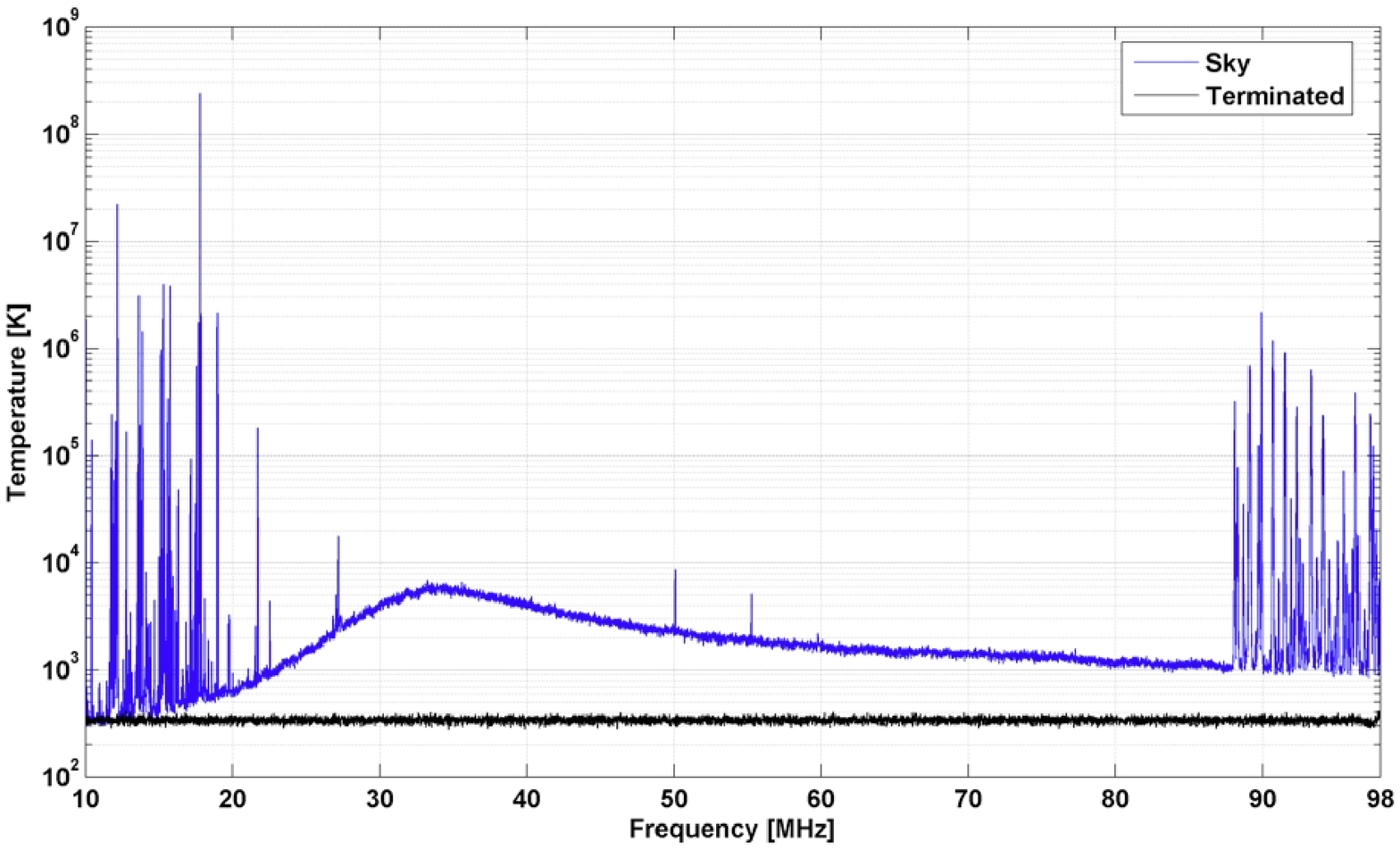}
\end{center}
\caption{A spectrum from the summed 7-element beamformer after 10 seconds
of integration with a spectral resolution of 6 kHz.   The receiver noise
temperature is about 250 K and roughly constant across the LWA
frequency range (black line).  The signal from the sky is shown 
in blue.  The turnover in the sky signal near 33 MHz is a result
of the decreasing efficiency of the LWA dipoles. }
\label{skydominance}
\end{figure}

To accommodate the various uncertainties in the RFI environment, we
have developed an ARX which can be electronically reconfigured between
three modes: A full-bandwidth (10-88~MHz) uniform-gain mode, a
full-bandwidth dual-gain mode in which frequencies below about 40~MHz
can be attenuated using a ``shelf filter,'' and a 28--54~MHz mode,
which serves as a last line of defense in the case where RFI above
and/or below this range is persistently linearity-limiting.  A 
sample spectrum, showing the strong RFI at the top and bottom 
ends of the spectrum, is shown in Figure~\ref{skydominance}.  The
noise is dominated by that from the sky, as can be seen by 
comparison of the sky signal with that of a terminated load.  

\bigskip

\begin{figure}
\begin{center}
\includegraphics[width=.8\textwidth]{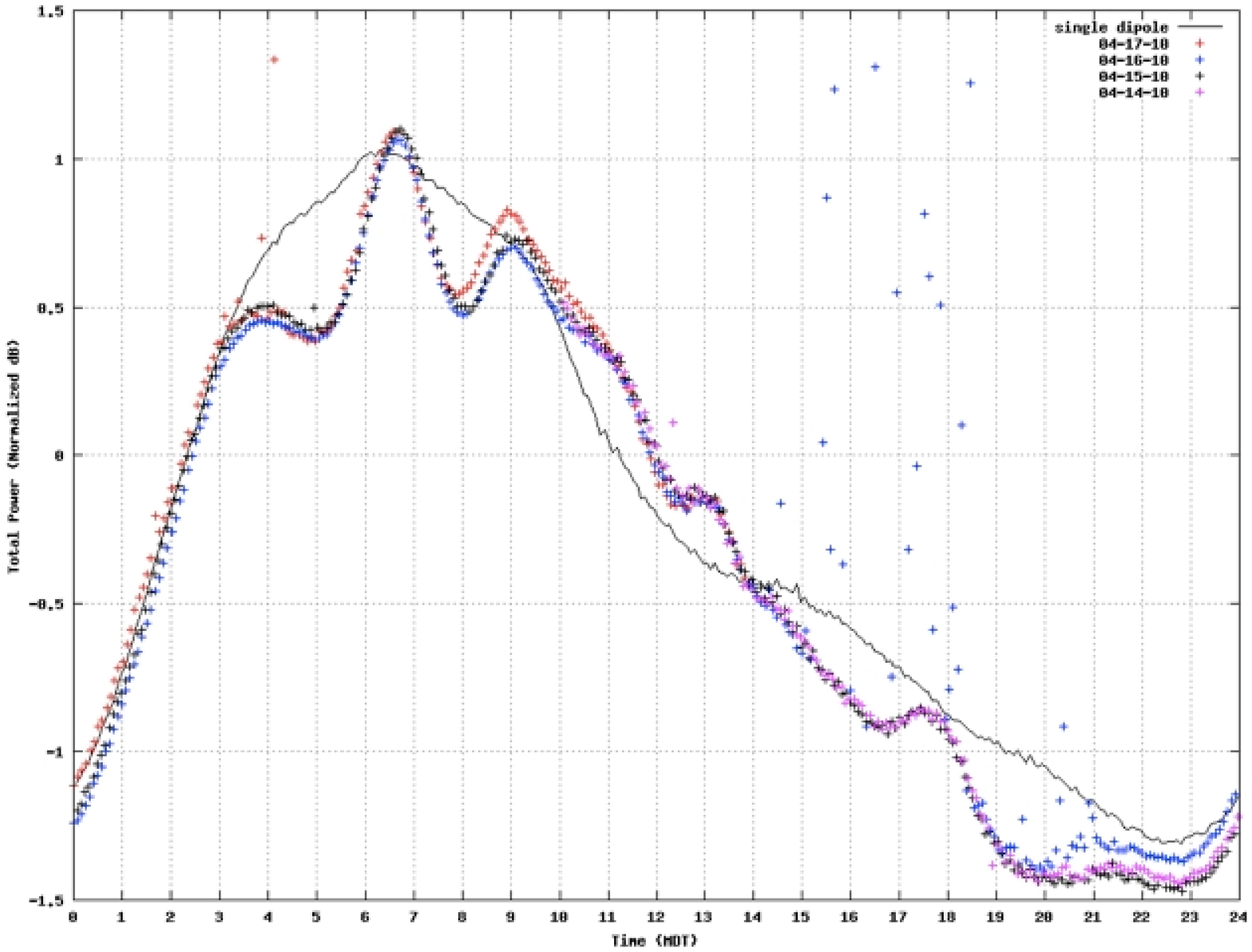}
\end{center}
\caption{Plot showing total power for a single dipole (thin black
  line), integrating for 2 seconds every 5 minutes and for the sum of
  7 dipoles (crosses) integrating in a similar fashion on three
  consecutive days.  The discrepant high points are most likely due
  to thunderstorm activity. The frequency employed was 72.25 MHz with a
  3 MHz bandwidth. }
\label{totalpower}
\end{figure}

The installation of the dipole antennas for LWA-1 is complete.  At the time
of writing (July 2010), cabling is on track for completion in August
2010.  The system is currently operating with an interim
16-antenna analog beamformer, using the Eight-meter-wavelength
Transient Array (ETA) ``S60'' digital receiver, and data capture
system.  In Figure~\ref{totalpower} we plot the total power received 
for a single dipole and for 7 dipoles summed together over the 
course of three days.  Some discrepant values are the result of
thunderstorm activity on the afternoon of April 15.  

\subsection{LWA-1 Sensitivity}

\begin{figure}
\begin{center}
\includegraphics[width=.8\textwidth]{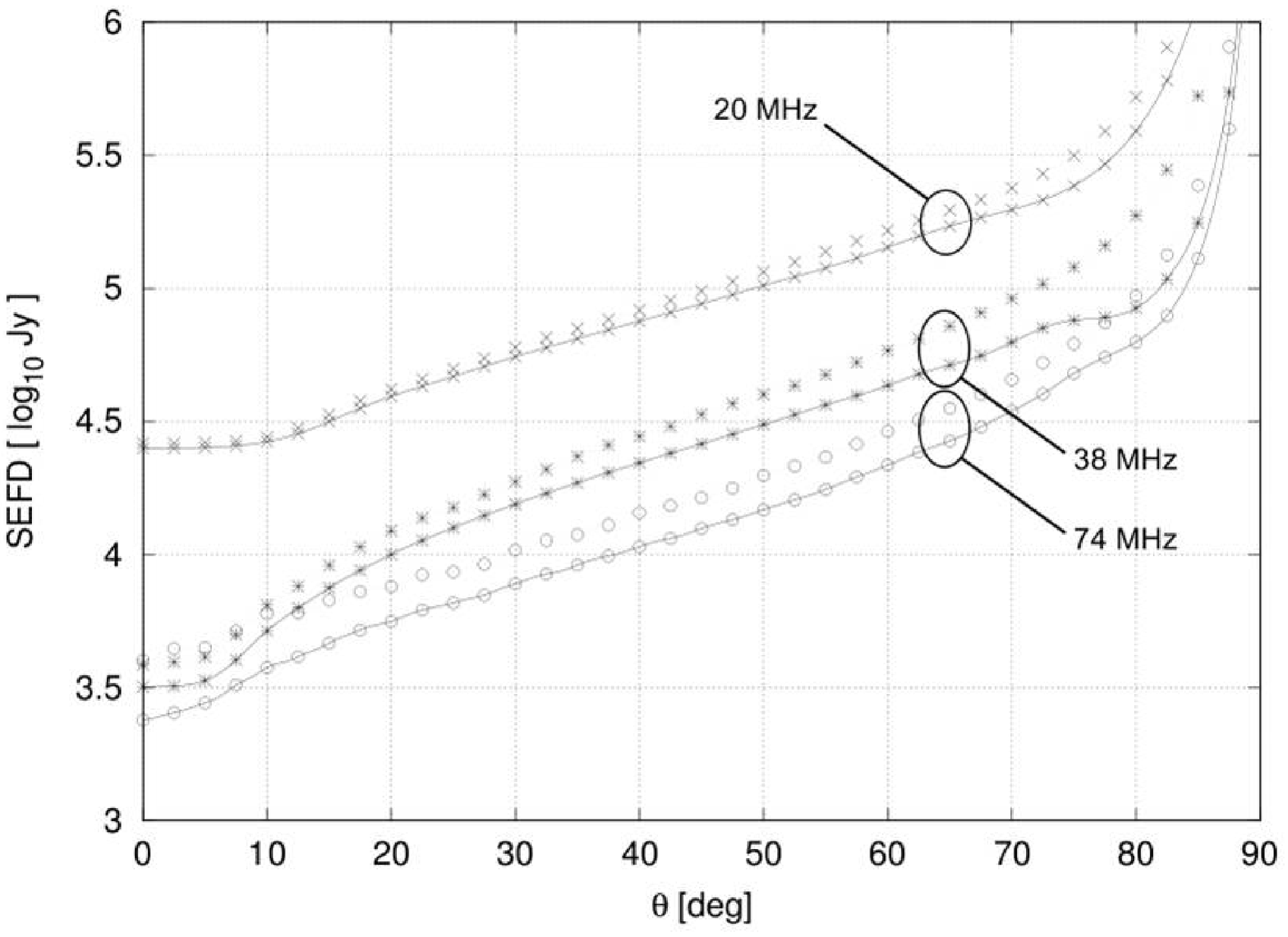}
\end{center}
\caption{Calculated SEFD of LWA-1 as a function of degrees away from 
the zenith.  For each frequency, 
the upper (dotted) curve is the result for simple
beamforming, and the lower (solid) curve is the result for optimal beamforming.
From \cite{LWA166}.}
\label{SEFD}
\end{figure}

At low frequencies, Galactic noise can be a significant or dominant
contribution to the total noise.  This, combined with mutual coupling
between antennas, makes it difficult to predict the sensitivity of
these instruments. Ellingson (2010)\cite{LWA166} describes a system
model and procedure for estimating the system equivalent flux density
(SEFD) -- a useful and meaningful metric of the sensitivity of a radio
telescope -- that accounts for these issues.  The method is applied to
LWA-1, and it is shown that the correlation of Galactic noise
between antennas significantly desensitizes the array for beam
pointings that are not close to the zenith.  It is also shown that
considerable improvement is possible using beamforming coefficients
that are designed to optimize signal-to-noise ratio under these
conditions (see Figure~\ref{SEFD}).  The receiver noise is about 
250 K, but has little influence on the SEFDs which range between 
3,000 and 100,000 Jy over the frequency and elevation range plotted.

\subsection{Radio Frequency Interference and Deep Integrations}

Radio Frequency Interference (RFI) is always an issue when working
at low frequencies.  We have sought to minimize internally-generated 
RFI by use of careful shielding of the station electronics.  For
externally-generated RFI we have chosen
modes and gain settings based on a detailed study
of RFI at the EVLA, combined with a study of A/D capabilities,
leading to the conclusion that an A/D of about 200~MSPS with 8-bit
sampling was probably sufficient when combined with an ARX having
the capabilities described above.  We currently favor
a sampling rate $F_s = 196$~MSPS, as this results in the highly
desirable situation that the 88-108~MHz FM broadcast band aliases
onto itself, which greatly reduces anti-alias filtering requirements.

\begin{figure}
\begin{center}
\includegraphics[width=.8\textwidth]{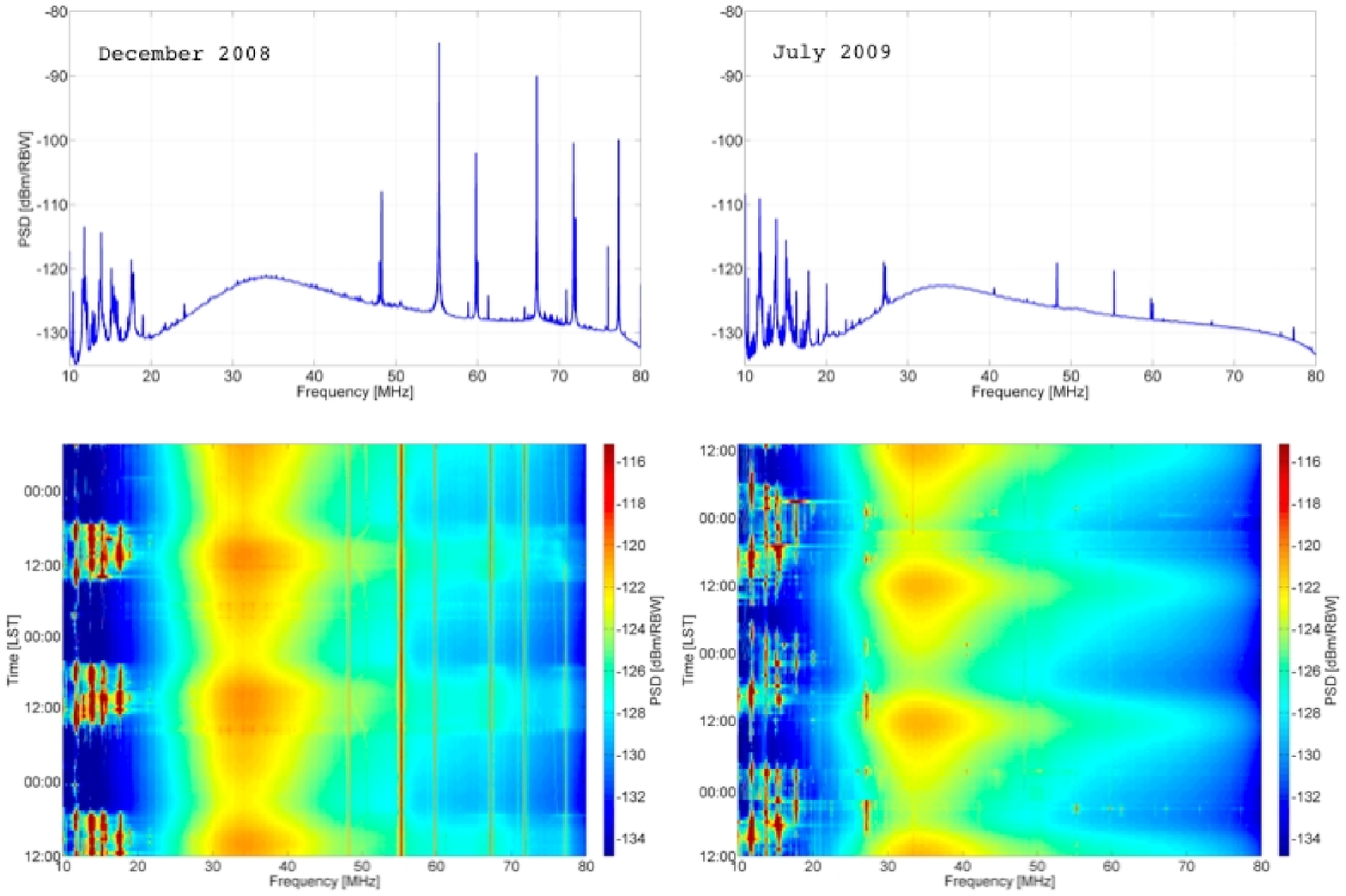}
\end{center}
\caption{RFI environment at LWA-1.  The left panels show an indicative
spectrum and waterfall plot
before the switch to digital TV.  The right panels show data taken after the strong 
analog TV stations have vacated the spectrum. The spectrum is
Galactic noise dominated. The diurnal variations seen in the waterfall plots
are predominantly due to the culmination of the Galactic Center, and day-night variation of HF noise.}
\label{RFI}
\end{figure}

During construction of the first station, the RFI situation has 
actually improved with the conversion from analog to digital 
television (Fig.~\ref{RFI}).  At the time of the conversion in 
July 2009, all of the television stations in New Mexico elected
to relocate their digital transmissions to frequencies above
the LWA band.

\begin{figure}
\begin{center}
\includegraphics[width=.9\textwidth]{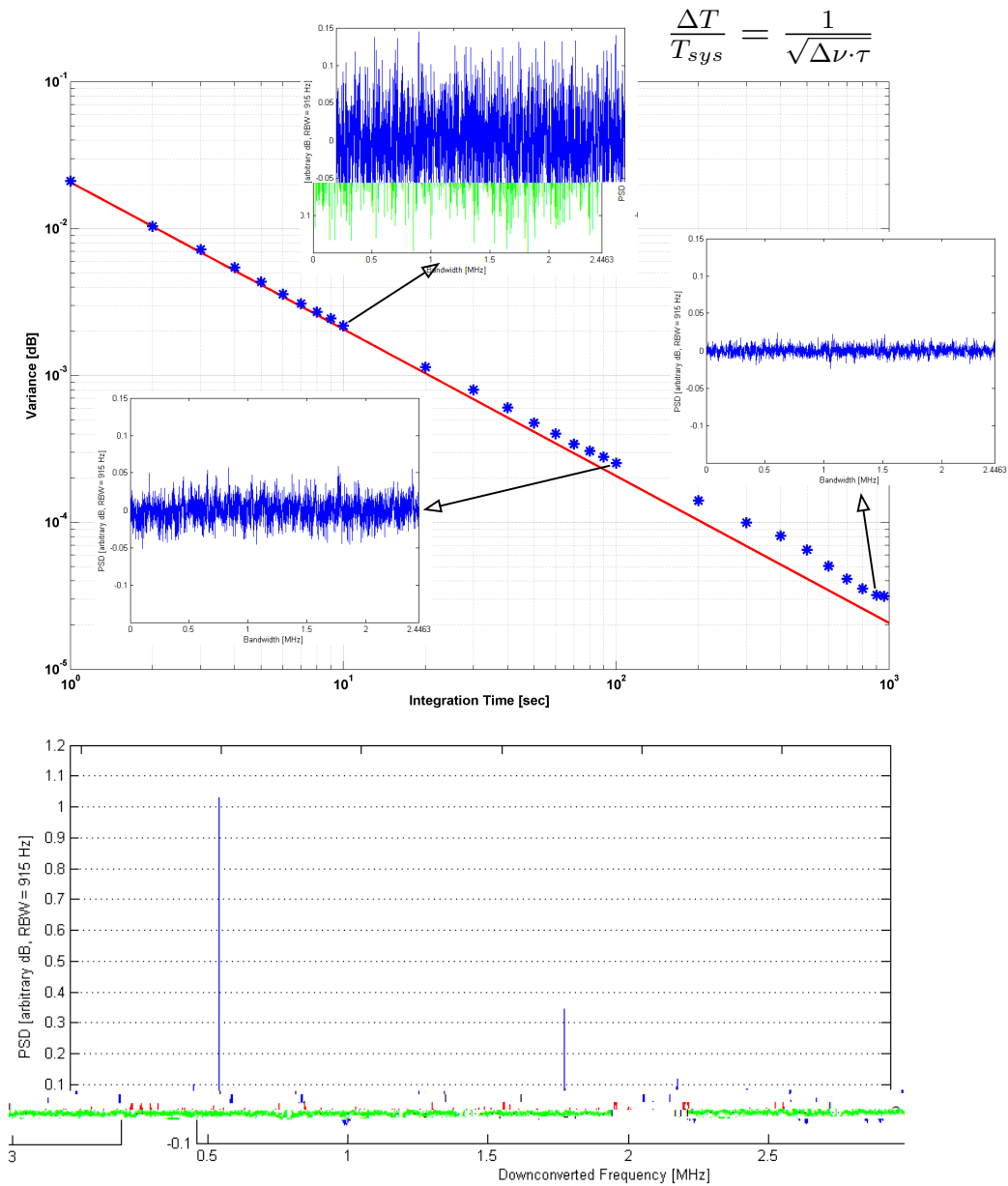}
\end{center}
\caption{Deep integration noise behavior.  The noise continues to drop as $1/\sqrt{t}$ for integrations
as long as an hour.  These drift scan data were taken at night, with 2.45 MHz bandpass (lower panel), 
centered at 72.25 MHz.  About 2600
channels are shown.  Data were corrected for diurnal total power variation.
(The step at about 100 sec is possibly due to HVAC turning on.) 
Total integration times where noise continues to drop radiometrically have exceeded an hour,
beyond the time indicated here.}
\label{noise}
\end{figure}

Strong RFI can cause a host of problems including compression in the 
receivers, and aliasing.  Dealing with strong RFI requires the
appropriate design choices, especially in the analog receiver, as
described previously. Weak RFI can be just as damaging if it masks the
faint cosmic signals that one is searching for.  While station
electronics are still being completed, deep integrations with a
portion of the array show that the noise behaves radiometrically
beyond an hour of data collecting (Fig.~\ref{noise}).  These results
were obtained at night, with 3 MHz bandwidth, centered near 74 MHz.
About 20\% of the frequency band was excised due to the presence of
low-level RFI in this test.  Deep integrations have paused while the
array is being finished, so the duration of the deepest integrations
may well be longer that the hour+ limit indicated by current tests.
This is indeed promising for the prospects of deep imaging.

\subsection{Digital Signal Processing}

The signal from every antenna is processed by a dedicated
direct-sampling receiver consisting of an analog receiver (ARX) and an
analog-to-digital converter (A/D) as described in \S2.1.  Beams are
formed using a time-domain delay-and-sum architecture, which allows
the entire 10--88~MHz passband associated with each antenna to be
processed as single wideband data stream.  Delays are implemented in
two stages: A coarse delay is implemented using a first-in first-out
(FIFO) buffer operating on the A/D output samples, followed by a
finite impulse response (FIR) filter, which is also used to introduce
corrections for polarization and other frequency-dependent effects.
The raw linear polarizations are transformed into
calibrated standard orthogonal circular polarizations, and
The signals are then added to the signals from other antennas
processed similarly.  Four dual-polarization beams of
bandwidth 78~MHz, each capable of fully-independent pointing over the
visible sky, will be constructed in this fashion.  

The beams will be available for various ``backends'' implemented at
the station level, such as data recorders, wideband spectrometers,
and pulsar machines. For interferometric imaging, two ``tunings''
will be extracted from any frequency in the 78~MHz-wide passband,
having bandwidth selectable between 400~kHz and 20~MHz divided into
4096 spectral channels.  This is the output to the LWA correlator.

\subsection{Prototype All Sky Imager}

The Prototype All Sky Imager (PASI) will consist of a software
correlator, and a near real time imager, both operating on an IBM
cluster already purchased for these tasks.  The PASI will be a
``back-end'' instrument designed specifically for the first station of
the LWA.  Together, this equipment will allow us to image 75\% of the
sky every 24 hours, with an instantaneous field-of-view of over
120$^\circ$ $\times$ 120$^\circ$.  This will provide unparalleled
resolution and sensitivity all-sky imaging in a largely unexplored
frequency band.  Expected sources to be imaged include the Sun,
planets, flaring stars, active galaxies, quasars, magnetars,
black holes, and gamma-ray bursts.  The discovery of entirely new 
classes of objects are possible and will be followed up by 
observations with the LWA and other ground (e.g., EVLA) 
and space-based (e.g., Chandra) facilities.  

\section{Timeline and Future Plans}

Currently the first LWA station is using analog beamformers and the
ETA ``S60'' digital receiver to commission the station.  During 
summer 2010 this system will be increased to make use of up to 32 antennas.
At the same time, the first digital processing board will be installed
with the capability of handling up to 20 antennas.  This will allow
for commissioning of the transient buffer modes (TBW and TBN)
initially, and later the beam outputs.  The installation of the 
complete analog and digital electronics will continue throughout
the fall resulting in a fully operational station in early 2011.
The first observing programs have already been selected and some
of them have even begun taking data.  

\begin{figure}
\begin{center}
\includegraphics[width=.9\textwidth]{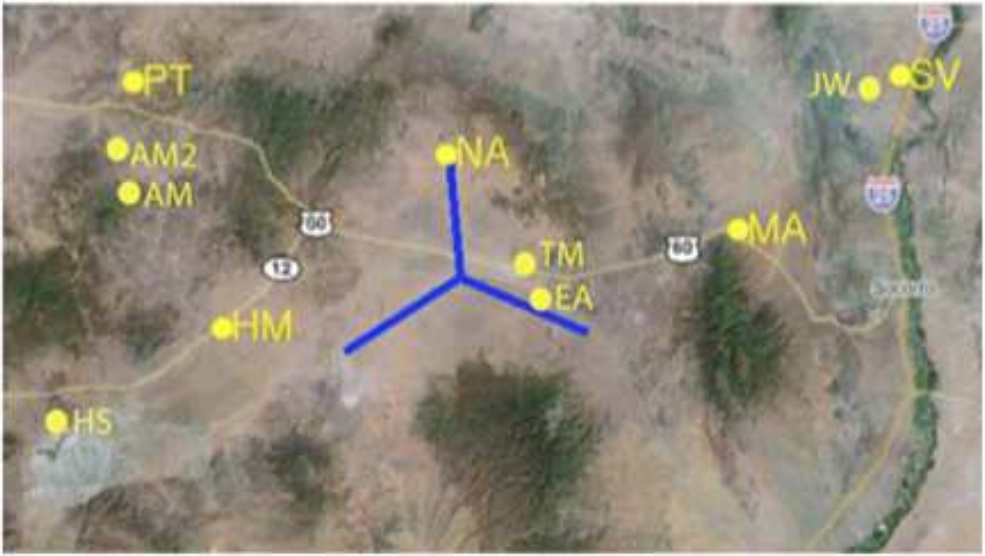}
\end{center}
\caption{Plot showing planned stations, including LWA-2 (North Arm) and LWA-3 
(Horse Mountain).}
\label{nextstations}
\end{figure}

Beyond the first LWA station we are actively seeking funding for
additional stations.  The land and infrastructure for a second LWA
station (labeled NA in Figure~\ref{nextstations}) is in place, 19 km
north of LWA1.  The land for a third station at Horse Mountain (HM) 
is also leased, but has not yet been developed.  Leases for two
other sites (marked MA and HS in Figure~\ref{nextstations}) are 
pending.

Basic research in radio astronomy at the Naval Research Laboratory is
supported by 6.1 base funding.

\end{document}